\documentclass[epj]{svjour}

\usepackage{epsfig}

\def\be{\begin{equation}}
\def\ee{\end{equation}}
\def\bea{\begin{eqnarray}}
\def\eea{\end{eqnarray}}

\begin{document}
\title{Stochastic equations generating continuous multiplicative cascades}

\author{Fran\c{c}ois Schmitt\inst{1} \and David Marsan\inst{2}
}                     
\institute{Vrije Universiteit Brussel, Dept. of Fluid Mechanics,
Pleinlaan 2, 1050 Brussels, Belgium,
\email{francois@stro.vub.ac.be}
\and
LGIT, Universit\'e de Savoie
Campus scientifique
73376 Le Bourget du Lac, Cedex, France
}
\date{Received: date / Revised version: date}
%
\abstract{
Discrete multiplicative turbulent cascades are described using
a formalism involving infinitely divisible random measures.
This permits to consider the continuous limit of a cascade developed on a 
continuum of scales, and
to provide the stochastic equations defining such processes,
involving infinitely divisible stochastic 
integrals. Causal evolution laws are also given. 
This gives the first general stochastic equations 
which generate continuous multifractal measures or processes.
} 
\maketitle

\section{Introduction}

Multiplicative cascades were first introduced 
in turbulence to model the energy flux in the inertial
range. The cascade formalism was originally introduced
as a discrete (in scale) procedure, with
a fixed (often 2) scale ratio between the scale
of a structure and that of the daughter structures
(see \cite{yagl66,mand74,sche84,benz84}).
Discrete cascade models lead to discrete scale invariance,
characterized by log-periodic modulations  
\cite{novi66,novi90,sorn98a,sorn98b}. On the other hand, a continuous
symmetry leading to scaling for any scale ratio
has been proposed, and corresponds to a scale densification of cascade models 
\cite{sche87,sait92,novi94,dubr94,she95,cast96}.
Scale densification implies the use of
infinitely divisible (ID) random variables, defining cascade models 
that can be called log-ID \cite{sait92,novi94,she95},
and have been compared to experimental
data in various studies 
\cite{schm92,she94,benz95,chil96,sche97,arne98}.
This then leads to an interrogation: discrete cascade models are built
using a simple recursive multiplicative procedure (see below), but
what is the continuous limit of this procedure?
What is the stochastic equation generating continuous
multifractals? Up to now, the process generated by
such continuous multiplicative cascades has not been 
explicitely described in the
general case; only its statistical moments are given \cite{she95},
or some general relation verified by the pdf at different scales \cite{cast90}.
This paper aims at detailing how ID random measures can
be introduced for
discrete cascades; their continuous limit is then given
in the form of ID stochastic integrals.
The stochastic evolution laws that generate
causal continuous multifractal processes will also be provided.
This is of direct importance for providing estimators of the future
state of the process.

\section{Discrete multiplicative cascades }

Soon after Kolmogorov and Obukhov published their
lognormal proposal for the statistics of the small-scale
dissipation field \cite{kolm62,obuk62},
experimental studies showed that the dissipation field
had long-range power-law correlations \cite{gurv63,pond65}.
This lead Yaglom to propose a random cascade model
with long-range correlations
and small-scale lognormal statistics \cite{yagl66}.
Yaglom's multiplicative cascade model is at the basis of
most cascade models introduced later to account for turbulent intermittency.
It is a discrete (in scale) model, but most
of its properties are shared by continuous models.
A lognormal pdf is assumed, but this is an unnecessary
hypothesis, as is now well recognised.
This model is multiplicative, nested in a recursive manner.
The multiplicative hypothesis generates large fluctuations,
and the stacking generates long-range correlations,
giving spatially to these large fluctuations their
intermittent character.

As is classically done (see e.g. Frisch \cite{fris95}), 
we define the cascade yielding a dissipation field
$\epsilon(x)$ at the smallest scale $\ell_0$, as the product 
\begin{equation}
\epsilon(x) = \prod_{i=1}^n W_{i,x}
\label{eq1}
\end{equation}
\noindent of $n$ independent realisations $W_{i,x}$
of a common, positive law. The cascade is developed from the largest scale
$L$ down to $\ell_0=L/\Lambda$ where $\Lambda=\lambda_1^n$ is the total
scale ratio and $\lambda_1>1$ is the constant scale 
ratio between two
consecutive scales. Generally one assume for convenience $\lambda_1=2$, but we will
later on consider the $\lambda_1\to 1$ limit corresponding to a continuous cascade
\cite{sche91}. Since all random variables are independent, one
has the moments of order $q>0$ of $\epsilon$:
\be
 < (\epsilon(x))^q > =  \prod_{i=1}^n < \left( W_{i,x}\right)^q > 
 = < W^q >^n = \Lambda^{K(q)}
\label{eq2}
\ee
where $K(q) = \log_{\lambda_1} < W^q >$. For discrete cascades, the analytical
expression taken by $K(q)$ is only loosely constrained a priori:
by conservation $K(0)=0$, $K(1)=0$,
and since $K(q)$ is -- up to a $\log \lambda_1$ factor -- the
second Laplace characteristic function of the random variable
$\log W$, it is a convex function (see \cite{fell71}).

To densify the cascade described above,
we keep the total scale ratio $\Lambda$ large but fixed;
the continuous limit can be obtained by increasing the total step
number $n$, hence $\lambda_1 = \Lambda^{1/n} \rightarrow 1^+$
(see \cite{sche91,she95,sche97,schm98}).
Equation (\ref{eq1}) then shows that, in this limit, 
$\log \epsilon$ is an ID random variable
(see \cite{fell71} for ID random variables):
continuous cascade models
are log-ID \cite{sait92,novi94,she95,cast96}. 
This restricts the eligible
cascade models, since ID laws define a specific family of
probability distributions. 

Let us also mention one of the main properties of 
multiplicative cascades: long-range power-law correlations.
Following the development given by Yaglom \cite{yagl66},
one can consider two points separated by a distance $r$
as having common ancestors from steps $1$ to $p$,
and separated (hence the corresponding random variables are
independent) paths for steps $p$ to $n$, where $\lambda_1^p \approx r$. 
Direct calculations then provides the classical result
for two-points correlations of multifractal fieds
\cite{cate87,mars96}:
\be
 <\epsilon(x)^p\epsilon(x+r)^q> \approx
  \Lambda^{K(p+q)}  
r^{K(p)+K(q)-K(p+q)}
\label{eq3}
\ee
for $p>0$ and $q>0$. Since $K(q)$ is non-linear for multifractal
distributions, the exponent $K(p)+K(q)-K(p+q)$ quantifies the long-range power law correlations
of multifractal measures. For $p=q=1$, this yields the
$\mu=K(2)$ exponent originally given for usual correlations
by Yaglom \cite{yagl66}.

The scaling law for the moments (\ref{eq1}) and the
power-law correlations (\ref{eq3}) are the two signatures
of multifractality, that are to be recovered by multifractal
stochastic models, as we define them below.

\section{ID random measures and stochastic integrals}

The densification of a multiplicative
cascade implies that $\log\epsilon$ is an ID random variable; we now express this
densification in the form of ID stochastic integrals.
Since we need below to consider moments of order $q>0$
of the ID random variable $\Gamma=\log \epsilon$,
we consider ID laws for which the second Laplace
characteristic function $\Psi_X(q) = \log < e^{qX} >$
converges for a given domain $\Theta$.
For an ID random variable $X$, one has the general result
that $\forall n$ integer,
$\frac{\Psi_X(q)}{n}$ is still a second characteristic function.
This shows that a family of ID laws can be defined: two ID laws
belong to the same family if their second characteristic function
is proportional. Then each ID family can be characterized by a 
reference function $\Psi_0(q)$. We choose a reference function
such that $\Psi_0(1) = 1$, and for an ID random variable $X$,
we define its scale $S(X)$ as the proportionality factor,
giving the general identity $\Psi_X(q) = S(X)\ \Psi_0(q)$.
The scale is a positive real number; it is an additive 
function since for two
independent ID random variables $X$ and $Y$ of the same family, 
it is easily checked that we have $S(X+Y) = S(X)+S(Y)$.
An ID random variable is then uniquely characterized by its reference
second characteristic function and its scale (relative to this reference function). 
As examples, $\Psi_0(q)=q^2$ for a Gaussian law,
and $\Psi_0(q)=\log_2(1+q)$ for a Gamma law.

We then define ID random measures as set functions
$M(A)$, such that $\forall A$, $M(A)$ is an ID random variable,
with a scale given by $S(M(A)) = m(A)$,
$m(A)$ being the control measure of $M(A)$. We can easily
check that $M(A)$ possesses the
basic additive property of random measures: for two sets
$A$ and $B$ with $A \cap B =\emptyset$, let us note  $C= A \cup B$.
By definition, $M(C)$ is a ID law, and its scale
verifies:  $S(M(C)) = m(C)$ $= m(A)+m(B)$ $=S(M(A))+S(M(B))$,
hence that  $M(C) = M(A)+M(B)$.
This corresponds to the following second characteristic function for $M(A)$: 
\be
          \Psi_{M(A)}(q) = m(A)\ \Psi_0(q)
\label{eq5}
\ee
This expression takes a more familiar form in the 1D case when
$A=[0,t]$ is an interval, and taking $M(A)=Y(t)-Y(0)=Y(t)$ where $Y$ is a 
process with independent and stationary increments. Then one has 
the classical result $\Psi_{Y(t)}(q) = t\ \Psi_Z(q)$
where $Z$ is the stationary process given by $Z(t)=Y(t)-Y(t-1)$. 
In the following we keep the
random measure notation, which is more general.

Having introduced an ID random measure $M$,
a stochastic integral can be built (see e.g. \cite{jani94}),
as a Stiltjes integral:
\bea
          \int_a^b f(t) M(dt) & = &
   \lim_{n \to \infty} \sum_{i=0}^{n-1} f( a+i\frac{b-a}{n})\nonumber\\
&& M ( [a+i\frac{b-a}{n},a+(i+1)\frac{b-a}{n}])
\label{eq6}
\eea
As we will see below, the densification of the cascade
leads to a stochastic integral with $f(t)=1$ $\forall t$. 
In this case, the second characteristic function of the
integral has a simple expression.
Let us note $I=\int_A M(dx)$. By additive property, $I$ is 
still an ID law of the same family as $M$,
and its scale is given by $S(I)=S\left(\int_A M(dx)\right)$
$=\int_A dx=m(A)$, such that we have still Eq. (\ref{eq5})
with $M(A)=\int_A M(dx)$ and $m(A)=\int_A dx$.
Let us note that when $f$ is not identically $1$, the result is not so
simple, since an addition of ID random variables belongs
 to the same family, but not a linear combination.
One has is this case (if the integral converges, see \cite{jani94}):
$ \Psi_{I}(q)= \int_A \Psi_0\left( qf(x)\right) dx$,
showing that in general, $\Psi_{I}(q)$ is not
proportional to $\Psi_{0}(q)$.

\section{Densification of the cascade and stochastic equations}

Let us introduce $\lambda$ a variable scale ratio, verifying
$1 \leq \lambda \leq \Lambda$, where $\Lambda$ is the fixed
largest scale ratio. We also introduce $R=\log \Lambda$
and $r=\log \lambda$. The elementary scale ratio
introduced above writes now $\lambda_1=\lambda^{1/n}=e^{R/n}$.
The discrete cascade corresponds to 
introducing a stochastic kernel $M$ and intervals $A_p$ and $B_p$ 
such that 
\be
\Gamma(x) = \log \epsilon(x)=\sum_{p=0}^{n-1}M \left(A_p, B_p(x) \right)
\label{eq8}
\ee
where here $M(A,B)$ is a random variable depending only
on $m(A)$, giving $\Psi_{M(A,B)}(q) = m(A) \Psi_0(q)$.
The intervals $A_p$ and $B_p$, responsible for the cascading
parent/children structure, are built in the following
way: the width of $A_p$ is linear in $r$, giving
$A_p = [ \frac{pR}{n},\frac{(p+1)R}{n}]$.
The intervals $B_p(x)$ are centered in $x$ 
and of width proportional to $\lambda_1^p = e^{pR/n}$, giving
$ B_p(x) = [x- \frac{\tau}{2}e^{pR/n},x+ \frac{\tau}{2}e^{pR/n}]$
where $\tau=L/\Lambda$ is the resolution. 
The densification, corresponding to $n \rightarrow \infty$,
transforms then Eq. (\ref{eq8}) into a stochastic integral
 and
using Eq. (\ref{eq6}) (with $f$=1), we obtain finally that:
\be
   \epsilon_{\Lambda}(x)=\Lambda^{-c} \exp  \int_{1}^{\Lambda}
M\left(\frac{cd\lambda}{\lambda},D_{\lambda}I_0(x)\right)
\label{eq9}
\ee
where $c>0$ is a parameter, 
$I_0(x)$ is the interval of length $\tau$
centered in $x$, and $D_{\lambda}$ is the dilatation operator
of factor $\lambda$. At a given position $x$, the 
stochastic integral corresponds to a kernel visiting a conical
structure, as represented in Fig. 2.
This expression can be generalized to
d-dimensional domains, and also to anisotropic scaling symmetries.

It can be easily verified that this stochastic equation
generates a multifractal field. Indeed, the moments are scaling as
$ < \epsilon_{\Lambda}^q >=\Lambda^{K(q)}$ with
\be
  K(q) = c\left(\Psi_0(q)-q\right)
\label{eq10}
\ee
Moreover the two-points statistics can also be
recovered: as was done above for Eq. (\ref{eq3}),
the correlation $< \left(\epsilon_{\Lambda}(x)\right)^p$
$\left(\epsilon_{\Lambda}(x+y)\right)^q >$
involves two integrals which have no intersection
(and thus are independent random variables) for
$\lambda < \lambda_0$, where $\lambda_0=e^{r_0}=\frac{y}{\tau}$.
$\lambda_0$ is the scale ratio of transition, as
shown in Fig. 3, and for $\lambda_0 \leq \lambda \leq \Lambda$,
the random variables corresponding to the two stochastic
integrals are no more independent. 
After some calculations, this leads to the same expression
as Eq. (\ref{eq3}) for discrete cascades.

We also give the expression for causal cascades, where
the position is time and the past does not depend
on the future. This case is of particular importance
for prediction of multifractal times series.
In this case, one can readily modify 
the intervals $B_p(t)$ by taking an
interval of the same length as before but 
preceding $t$: 
$B_p(t)=[t-\tau \exp(pR/n), t]$. 
This gives the following causal stochastic evolution
law for continuous multifractals: 
\be
   \epsilon_{\Lambda}(t)=\Lambda^{-c} \exp  \int_{1}^{\Lambda}
M\left(\frac{cd\lambda}{\lambda},[t-\tau \lambda,t] \right)
\label{eq11}
\ee

Let us finally consider an important family,
corresponding to logstable multifractals \cite{sche87,kida91},
including the lognormal case.
Stable laws are ID, and possess
a stronger property corresponding to stability,
which can be written here 
$M(kA) \doteq k^{1/\alpha} M(A)$ for $k>0$,
where $0 \leq \alpha \leq 2$ is the L\'evy index;
$\alpha=2$ for the Gaussian case \cite{jani94,samo94}. 
We have here $\Psi_0(q)=q^{\alpha}$; we note that
when $\alpha<2$, the second Laplace characteristic function
is defined for positive moments 
only for asymmetric laws for which hyperbolic pdf corresponds
to negative fluctuations ($\Pr(-X>x)\approx x^{-\alpha}$),
whereas positive fluctuations have an exponential decay
\cite{sche87,kida91}.
Then, by splitting
Eq. (\ref{eq8}) into two integrals, corresponding to
backward and forward domains, and introducing  
the change of variables $u=x-\frac{\tau}{2}\lambda$
and $v=x+ \frac{\tau}{2}\lambda$ respectively, 
one obtains (introducing the L\'evy measure $L_{\alpha}(du)=M(du,[u,x])$)
a stable stochastic integral:
\be
   \epsilon_{\Lambda}(x)=\Lambda^{-c} \exp  
\int_{A(x)}
\vert u-x \vert^{-1/\alpha} dL_{\alpha}(cu)
\label{eq12}
\ee
where $A(x)=[x-X/2,x-\tau/2] \cup [x+\tau/2,x+X/2]$.
This equation corresponds to the exponential of a fractional
integration (over a limited domain) of order $1-1/\alpha$
of a L\'evy-stable noise. 
This expression was already given \cite{sche87,mars96}
by phenomenological arguments, and is here derived as
a direct consequence of the densification.
When the position is time, we obtain the following causal
stochastic evolution equation for a logstable multifractal
generated with a fixed scale ratio $\Lambda=T/\tau$:
\be
   \epsilon_{\Lambda}(t)=\Lambda^{-c} \exp  
\int_{t-T}^{t-\tau} (t-u)^{-1/\alpha} dL_{\alpha}(c u)
\label{eq13}
\ee
This can be directly used in numerical simulations.
For lognormal multifractals $L_{\alpha}$ is replaced
by the Wiener measure $W$.

\section{Conclusion}
We now discuss the new results provided by our approach, 
compared to related papers. She and Waymire \cite{she95} 
have given a general expression for $K(q)$ for continuous
cascades, using the canonical L\'evy-Khinchine representation
for the second characteristic function of ID laws. This could of
course also be provided here. On the other hand, we have pushed further
the analysis, since the process itself was
not studied in \cite{she95}. Castaing and collaborators
have studied the convolution properties of the probability
density of velocity increments under a change of scale, for
continuous cascades \cite{cast90,cast96,chil96}. 
This property is recovered here (it can be 
obtained from Eq. (\ref{eq9})), and we also provide
the stochastic equation for the process itself.
Other studies have provided Fokker-Planck \cite{frie97}
or Langevin \cite{marc99}
equations for the cascade process in the scale-ratio space.
These equations apply only to lognormal cascades.
These studies consider a fixed spatial (or temporal) position 
and provide stochastic equations for the cascade process
developing at this particular position. 
This framework is then different from
ours by the fact that (i) the position is not taken into account,
whereas we included the position in our equations, and (ii) 
we provided equations for the general log-ID case, and not only for the 
lognormal case.

We have obtained continuous multifractals as the exponential
of a stochastic integral. This formalisation is of great
interest for theoretical studies since it provides the first general
stochastic equations for continuous multifractals. These equations
can be generalized to anisotropic situations, and their dynamical
properties (i.e. error growth and the corresponding predictability
limit, or return times) are now open to theoretical studies, whereas before
such studies were possible only through numerical simulations.
We have also provided the general evolution equation
for causal continuous multifractal processes. We have 
shown how these equations simplify for log-stable and lognormal multifractals.

These equations can be used for numerical simulations of
continuous multifractals. In the general log-ID case,
Eq. (\ref{eq9}) can be used as follows: a 2D ID noise
must be generated, in the $r$ ($r = \log \lambda$ is the logarithm
of the scale ratio) - position plane. Then for each position,
a numerical path-integration
is performed in this plane, as given by this equation. In the stable
case the procedure is simpler, since it is enough to simulate
a 1D stable (or Gaussian) noise and to proceed to a fractional
integration over this noise, as given by Eq. (\ref{eq12}).

\begin{acknowledgement}
The impetus to undertake this work has been given some time
ago by C. Nicolis, which is gratefully acknowledged.
The authors thank also S. Lovejoy, T. Over, D. Schertzer, 
S. Vannitsem for
useful discussions. An anonymous referee is thanked for
useful comments.
\end{acknowledgement}

\begin{figure*}[ht]
  \psfig{figure=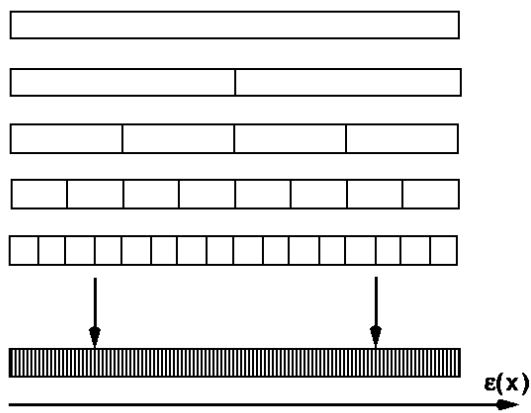,height=6. cm}
  \vspace*{-0.2 cm}
  \caption{{Schematic representation of a 
  discrete multiplicative cascade}
 \label{fig1}}
\end{figure*}

\begin{figure*}[ht]
  \psfig{figure=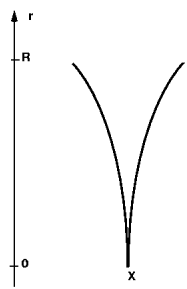,height=7. cm}
\vspace*{-0.2 cm}
  \caption{{Conic structure starting at position 
$x$, corresponding to the integration surface to obtain $\Gamma(x)$.}
 \label{fig2}}
\end{figure*}
 
\begin{figure*}[ht]
  \psfig{figure=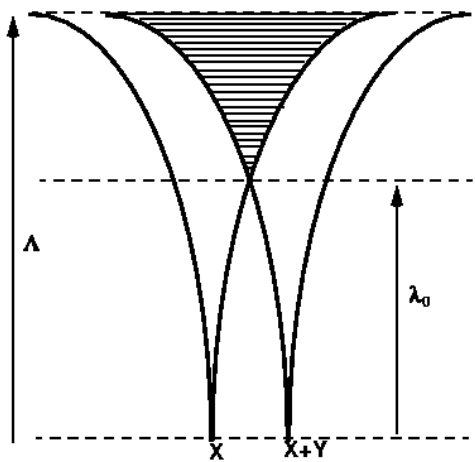,height=6.5 cm}
\vspace*{-0.2 cm}
  \caption{{Intersection of two conic structures centered at
  positions  $x$ and $x+y$.}
 \label{fig3}}
\end{figure*}

%
%
%
%

\end{document}